# Modelling and evaluation of a multi-tag LED-ID platform

Grzegorz Blinowski
Institute of Computer Science,
Warsaw University of Technology,
Nowowiejska 15/19, 00-665
Warszawa; Poland
Email: g.blinowski@ii.pw.edu.pl

Adrianna Kmieciak
Institute of Computer Science,
Warsaw University of Technology,
Nowowiejska 15/19, 00-665
Warszawa; Poland
Email: akmiecia@elka.pw.edu.pl

*Abstract*— **An LED-ID system works like an electronic "tag" transmitting a short digital broadcasted message. Low complexity LED-ID installations, being a subset of an emerging class of visible light communication (VLC) systems, may be considered as a replacement of popular RFID tags, Bluetooth tags and Wi-Fi beacons. In this work, we focus on multi LED-ID environments with "dense" tag placement. The problems that we focus on are estimating the level of cross-tag interference and the issue of tag proximity: how closely can we place the tags without making the system unusable? We present a theoretical model with a numerical simulation of sample arrangements. We also describe the results of experiments we conducted in a real-world test environment under different external lighting conditions.**

## I. Introduction

Visible light communication (VLC) is wireless optical communication technology through which baseband signals are modulated on the light emitted by an LED [1] – [3]. The decreasing cost and hence rapid adaptation of LED-based light make VLC a promising communication technique and an excellent alternative to radio-based wireless communication. A unique feature of a VLC system is that it performs two functions simultaneously: illumination and communication. This results in a reduction of costs because a separate system for data transmission is not needed any more – existing illumination infrastructure is used instead.

VLC systems have been proposed and implemented both for indoor and outdoor applications (see [2] and [4]). Indoor applications include a range of communication facilities provided today by WLAN and personal area networks (PAN) such as office communication [5], multimedia conferencing [6], peer-to-peer data exchange, data broadcasting (especially multimedia such as home-audio and video streams – see [7] – [10]). A relatively simple VLC system is able to achieve data rates of up to 100 Mbit/s over a distance of 1 – 3 m with a single light source and a simple equalized receiver [11]. Data rates of over 1 Gbit/s have been recently obtained for more complex transmitter-receiver configurations.

One application of VLC are LED-ID platforms, which can be used in numerous environments including shops and supermarkets, museums, plenum spaces, etc. An LED-ID system works as an electronic "tag" transmitting a short digital broadcasted message. LED-ID systems, with their low complexity, may be considered as replacements of popular RFID and Bluetooth tags. An example of LED-ID systems in use are "smart" supermarket carts, which via illumination infrastructure record a shoppers' path for subsequent analysis. LED-ID systems may also be used to "tag" particular shop shelves and areas to enable fast product localization. Digital signage systems used in museums, exhibitions, etc. are another example of LED-ID technology. These signage systems may be used with specialized applications for mobile platforms to provide information about objects in proximity. Yet another LED-ID field of application arises in environments where the usage of radio-based technology, such as Bluetooth, ZigBee or RFID, is hazardous or limited by regulations, for example in mines, petrochemical plants, aeronautics and hospitals.

In comparison to more complex VLC systems, LED-ID tags are simple: their functionality is limited to broadcasting digital information. LED-ID tags typically do not provide duplex communication; tag "programming" is done via wired or wireless connections and in some cases the ID is simply hardcoded into the tag's microcontroller unit. In many cases, the tag is simple enough that it does not support cooperation in a multi-transmitter environment – it simply broadcasts its information with no regard for other tags competing for the same medium. As was explained in [12] and [13], an optical communication link can be modelled as a Poisson channel. In the general case of multiple transmitters, it was shown that the maximum total throughput of the Poisson MAC monotonically increases with the

---

This work was supported by the Statutory Grant of the Polish Ministry of Science and Higher Education to the Institute of Computer Science, Warsaw University of Technology.

number of transmitters and is bounded from above. Therefore, adding more inputs to a Poisson MAC eventually saturates the entropy rate (and hence the information content) of the output. Given the channel capacity limitation, a signal source with sufficient transmitting power will be able to saturate the channel, obscuring the data source. The same result may also be obtained by a larger number of low-power transmitters.

In this work, we will focus on multi LED-ID environments with "dense" VLC tag placement. Examples of such environments include article tagging on shop shelves, the tagging of individual items in museum exhibitions, and other cases where light-tagged items are placed closely together. In such environments with dense arrangements of tags, the cones of light emitted by different luminaires overlap. The problems that we focus on in this work are as follows: what measures may we use to evaluate such an environment? What is the level of cross-tag interference? How closely can we place the tags without making the system unusable?

The structure of this paper is as follows: in section II we present the architecture of LED-ID systems, which leads us to the theoretical system model then described in section III. We use the model for the numerical simulation presented in section IV. In section V, we show the results of an experiment that we conducted on a sample installation built from commercially available LED-ID components. Our work is summarized in section VI.

## II. ARCHITECTURE OF AN LED-ID SYSTEM

An LED-ID system consists of a transmitter ("tag") and a receiver ("reader"). The transmitter must be able to modulate the emitted light to transmit the digital tag. It consists of a luminaire which may use one or more LEDs (typically a high power white-light LED in blue-LED / yellow phosphorous technology), an LED-driver IC and a microcontroller unit driving the amplifier.

The critical difference between VLC and radio-based communication is that in VLC, data can not be encoded in the phase of the light signal. The information has to be encoded in the varying intensity of the emitted light. The demodulation depends on direct detection at the receiver - hence IM/DD (Intensity Modulated/Direct Detection) modulation techniques are used in VLC. Modulation in VLC must also take into account the requirements of dimming and flicker mitigation. Various modulation schemes have been proposed for VLC systems, including:

- On-Off Keying (OOK) - the data bits 1 and 0 are transmitted by turning the LED on and off respectively. In the "0" state, the LED is not completely turned off but rather the light intensity is reduced. The advantages of OOK include its simplicity and ease of implementation.
- Pulse Width Modulation (PWM) - the widths of the pulses are adjusted based on the desired level of light dimming while the pulses themselves carry the modulated signal in the form of a square wave.
- Pulse Position Modulation (PPM) - the position of the pulse in a series of pre-defined time-slots identifies the transmitted symbol.
- Orthogonal Frequency Division Multiplexing (OFDM) - the channel is divided into multiple orthogonal subcarriers and data is sent in parallel sub-streams modulated over the subcarriers. Standard "radio-based" OFDM techniques need to be adapted for application in IM/DD techniques because OFDM generates complex-valued bipolar signals which need to be converted to real values.
- Frequency Shift Keying (FSK) – the instantaneous frequency of a constant-amplitude carrier signal is changed between two (for BFSK) or more (for MFSK) values by the baseband digital message signal.

The modulation methods described above have numerous pros and cons [14]. OFDM is very effective in high speed transmission, when inter-symbol interference and multipath fading start to dominate the channel capacity. However, it is difficult to implement OFDM with the LED-driving analogue hardware that is currently used. PWM, PPM and their numerous variants provide light dimming and a simple way to eliminate flicker while maintaining good channel bandwidth. In some cases, the dominant factor in choosing a modulation method is the hardware available and its limitations. For example, with customer mobile devices, a plug-in photodetector is the simplest and the cheapest choice (see the receiver section below), and a compatible modulation method therefore must be used – FSK in this case. In this study, we assume that FSK modulation is used, as it is currently the dominant modulation method for mobile platforms.

In general, VLC systems may use two types of receivers: (1) a photodetector – typically a photodiode (a non-imaging receiver); (2) an imaging sensor (a camera). In LED-ID systems, where low cost is an important factor, simple photodetector receivers are used. Even with no or with very simple analog equalization they provide bandwidth that is more than adequate for LED-ID applications. In customer-grade VLC, a smart-phone or a similar device is used as a reader. In this case, the phone's built-in camera could be considered as the receiving device. However, this type of imaging sensor is very slow and inadequate for data transmission applications[1], hence plug-in photodetector modules compatible with a standard audio-in/out port are used instead.

### III. SYSTEM MODEL

The components of an LED-ID system include an LED transmitter consisting of one light source and a photodiode receiver. The received signal depends on the physical characteristics of the transmitting LED, the receiver, and channel characteristics. We use ray optics theory to calculate signal and noise levels and derive adequate metrics. We assume the Multiple Input Single Output (MISO) model, with multiple transmitting LEDs and one photodiode detector. A single transmitting LED is characterized by a half-power semi-angle and central luminous intensity (measured in candelas). The receiver is a simple non-imaging photodetector with an optical filter, optical concentrator and a single photodiode element with a field of view (FOV) angle, gain, a photodetector area and conversion efficiency (measured in A/W).

The metric that we use to measure the impact of the interference is bit error rate (BER), which depends on the signal to noise ratio (SNR) and modulation scheme. The relationship between BER and SNR depends on the modulation type and modulation parameters. For binary frequency shift keying (BFSK) with non-coherent detection [15]:

$$BER_{BFSK}(SNR) = \frac{1}{2}exp\left(-\frac{SNR}{2}\right) \quad (1)$$

we calculate SNR as follows:

$$SNR_s = \frac{\overline{s_{data}^2}}{\left(N + \overline{s_{interf}^2}\right)} \quad (2)$$

where $s_{data}$ is the data signal, $s_{interf}$ is the signal transmitted by other luminaires, and $N$ is noise.

The problem of noise in VLC environments has been studied in detail [16]. In general, the following noise sources should be considered: background and transmitter LED shot noise, thermal noise in the detector and the influence of inter symbol interference (ISI). The background or ambient noise comes from the sun and artificial light sources:

$$N = \sigma_{shot}^2 + \sigma_{thermal}^2 + \sigma_{ISI}^2 \quad (3)$$

where $N$ is the total noise variance and $\sigma_{shot}, \sigma_{thermal}, \sigma_{ISI}$ is the standard variance of shot noise, thermal noise and ISI respectively. The proper estimation of noise in VLC environments is crucial in studying the maximum attainable transfer rates under various conditions and modulation schemes. The input referred noise variance depends on the signal data rate. For low data rates in the range of $10^2 - 10^4$ bits/s, the major noise factor is shot noise:

$$\sigma_{shot}^2 = 2qRPB + 2qI_{bg}I_2B \quad (4)$$

Where $q$ is the electronic charge, $R$ is the responsivity of the photodiode, $B$ is the equivalent noise bandwidth, $P$ is the received power, $I_{bg}$ is the background current, and for a p-i-n/FET receiver we assume $I_2 = 0.56$. In the multi-luminaire study that we conduct in this paper, the dominant noise factor is the interfering signal from neighboring luminaires and not physical noise itself.

Now we will present the analytical model of the optical wireless channel which will let us derive SNR and BER measures for different physical scenarios. Our analysis is based on the fundamental paper by Komine and Nakagawa [17].

A single LED is a Lambertian emitter – its radiation intensity is a cosine function of the viewing angle and is given by

$$I(\theta) = P_t \frac{(m+1)}{2\pi} cos^m(\theta) \quad (5)$$

where $\theta$ is the irradiance angle, $P_t$ is the transmitted power and $m$ is the order of Lambertian emission given by irradiance semi-angle $\theta_{1/2}$ (half power angle)

$$m = -\frac{\ln 2}{\ln(\cos(\theta_{1/2}))} \quad (5)$$

Light propagates from the LED to the receiver via a channel which is modeled by direct channel transfer function $h_d$:

---

[1] It is possible to use a more complex multi-light source transmitter which takes advantage of the "imaging" properties of the sensor, however this is much more expensive than a simple single luminaire solution.

$$h_d = \begin{cases} \dfrac{(m+1)A\cos^m(\theta)}{2\pi d^2}\cos(\psi)\,R(\psi) & 0 \le \theta \le \theta_{FOV} \\ 0 & \theta > \theta_{FOV} \end{cases} \quad (7)$$

where $\theta$ is the irradiance angle, $\psi$ is the angle of incidence, $A$ is the receiver area, $R(\psi)$ is receiver gain, $d$ is the distance from the LED to the receiver and $\theta_{FOV}$ is the receiver's FOV semi-angle. The geometric model of this simple line of sight (LOS) case is shown in Fig. 1.

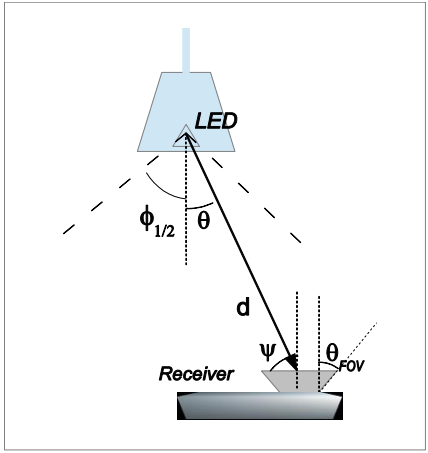

Fig. 1. Geometric model of LOS communication

For a single source, the output signal of the LED transmitter is given by the following general formula:

$$p_O(t) = P_t[1 + \mu\, x(t)] \quad (8)$$

where $P_t$ is the power transmitted from a single LED, $\mu$ is the modulation index and $x(t)$ is the modulating signal. Assuming that the receiver is DC blocked, we get the following general formula for the received signal:

$$s(t) = h_d\, P_t\, \mu\, x(t) \quad (9)$$

Considering the "legitimate" and "interfering" sets of transmitters, we obtain the following:

$$s_{data}(t) = \sum_{data\_LEDs} \{P_{LED}\, \mu\, x(t)\, h_d\} \quad (10)$$

$$s_{interf}(t) = \sum_{interf\_LEDs} \{P_{LED}\, \mu\, x(t)\, h_d\} \quad (11)$$

We use (10) and (11) in a numerical model to calculate BER as given in (1) for our study.

## IV. SIMULATION RESULTS

For our numerical simulations, we designed sample scenarios with 3 and 9 luminaires. The scenarios' dimensions are 2m x 2 m x 2m. We assume that the detector's photodiode is parallel to the luminaire plane. We simulated two luminaire placement scenarios: L1 - with 3 luminaires arranged in a line as shown in Fig. 2, and scenario G1 - with 9 luminaires arranged in a 3x3 square grid as shown in Fig. 3. The first scenario relates to a "shop shelf" arrangement and the second to an exhibition cabinet or stand. The physical parameters are summarized in Table I.

TABLE I
PHYSICAL PARAMETERS OF THE SIMULATED SCENARIOS

| Photodetector parameters | |
|---|---|
| FOV (field of view) | 60° |
| Detector area | 1 cm² |
| Detector gain | 1.3 |
| **Scenario parameters** | |
| Dimensions | 2m x 2 m x 2 m |
| Luminaire spacing Δx, Δy | L1: 16 cm  G1: 16 cm, 16 cm |
| # of luminaires, scenario L1, G2 | 3, 9 |
| **Luminaire parameters** | |
| Optical power | 1 W |
| Radiation semi-angle | 20° |

In both scenarios we show the logarithmic plots of the computed BER for data transmission. We assume that the BER level of maximum $10^{-2}$ is required for effective transmission of the LED-ID tag.

In scenario L1 we calculated BER for outer lamps, while the inner lamp is the interfering transmitter. BER is calculated on a plane at a distance of 30, 40 and 50 cm from the luminaire plane – Fig. 4. BER decreases as we move the receiver away from the luminaires and achieves values in the range of $10^{-6}$, $10^{-2}$ and $10^{-1}$ respectively. We can conclude that BER becomes intolerably high when the light cones (as limited by the radiation semi-angle) start to fully overlap each other, i.e. when the radius of the luminaire light cones is equal to the distance of their centers.

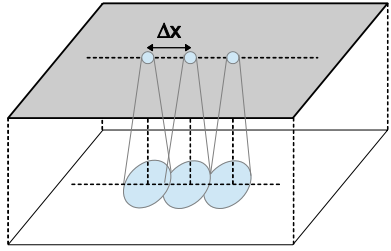

Fig. 2. Simulated scenario arrangement L1.

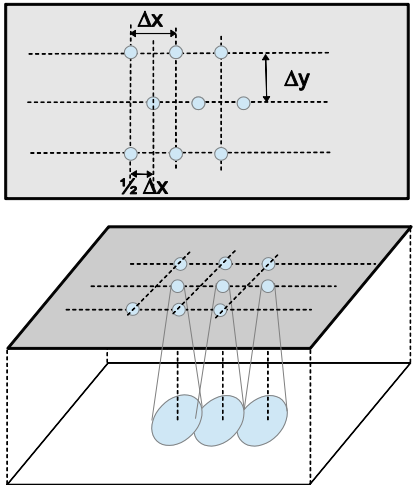

Fig. 3. Simulated scenario arrangement G1.

In scenario G1 we also calculated BER on a plane at a distance of 30, 40 and 50 cm from the luminaire plane – Fig. 5. BER decreases as we move the receiver away from the luminaires and achieves values in the range of $10^{-6}$ $10^{-2}$ and $10^{-1}$ respectively. The LED-ID tag under respective luminaires can be properly resolved, as was in the case of a single luminaire line.

The scenarios prove that the resolution of LED-ID tagging is quite satisfactory – even with dense luminaire placement, we are still able to obtain a reliable tag readout.

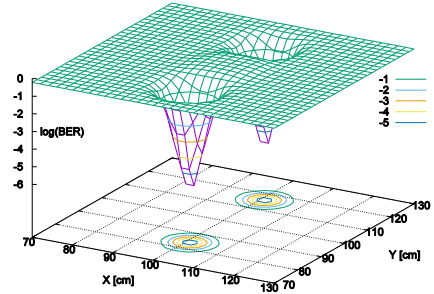

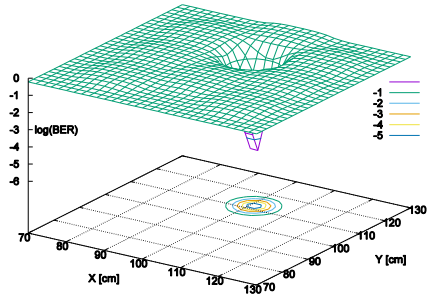

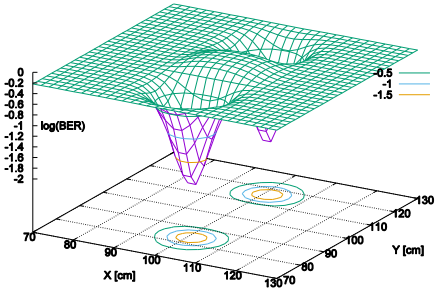

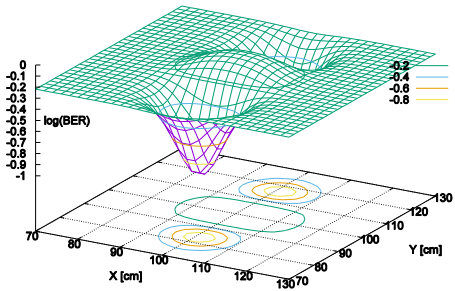

Fig. 4. BER - simulation results for scenario L1. From top: (1) outer luminaires, distance 30 cm; (2) inner luminaire, distance 30 cm; (3) outer luminaires, distance 40cm; (4) outer luminaires, distance 50cm.

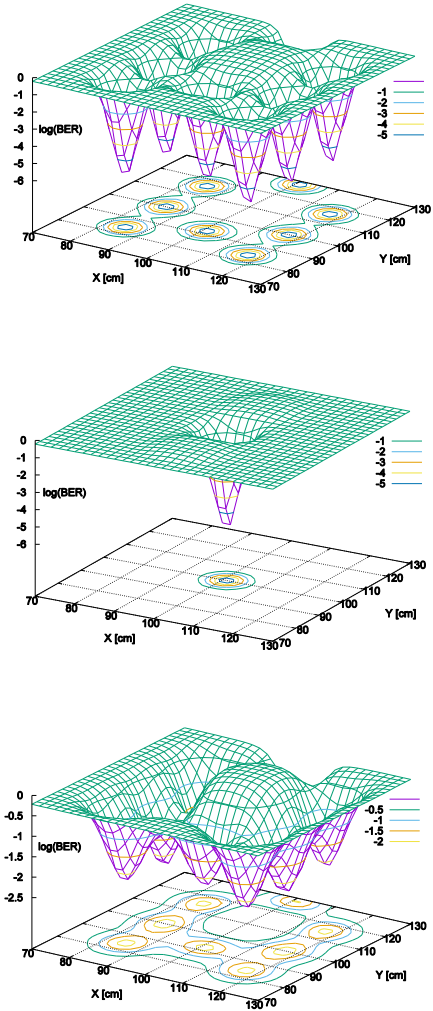

Fig. 5. BER - simulation results for scenario G1. From top: (1) outer luminaires, distance 30 cm; (2) inner luminaire, distance 30 cm; (3) outer luminaires, distance 40cm.

## V. SAMPLE SYSTEM EVALUATION RESULTS

For our experiments, we used LED-ID devices manufactured by OLEDCOMM as shown in Fig. 6. These luminaires came in the form of a desktop lamp with a 1W single LED light source, with a ~15° radiation semi-angle (as declared by the manufacturer, this parameter varies from unit to unit, and in most cases is a few degrees larger than declared). The luminous flux when measured 50 cm from the light source is ~ 900 lx (it varies by 5% between different luminaires). The OLEDCOMM kit also contained an audio-port plugin receiver compatible with most Android devices and an SDK library. The receiver uses a simple PIN photodiode with no optical concentrator or filter.

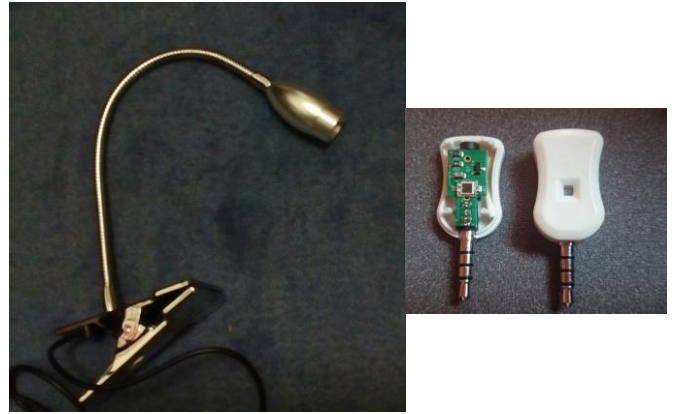

Fig. 6. LED-ID equipment used in experiments

For the tests, we implemented a client-server test suite consisting of an Android client program written in Java which gathers information regarding light intensity and lamp-ID numeric tags as reported by the library and sends it to the data-collecting server. The client has provisions for recording semi-automatic LED-sensor distance and is also able to buffer the data if the server is not accessible. Collected data may be manually tagged in the application to record various field conditions such as test series name, external illumination conditions, etc. The server stores data received for the client for further analysis. The server was implemented with the Django Rest Framework [18].

### A. Testing under various field conditions

To establish the baselines, we tested three sets of communication kits under the following external light conditions: (1) minimal external light source (< 10 lx); (2) ambient dispersed light (50 - 200 lx); (3) unmodulated direct light from an external LED source (up to 3000 lx); (4) direct sunlight (3000 – 5000 lx). The ambient light intensity levels were measured with a certified lux meter.

In each case we measured the maximum distance that guaranteed reliable ID transmission (5 tags correctly received in sequence). Measurements were collected with 1 - 5 cm intervals for $d$ and $x$ values – see Fig. 7, for 3 different luminaires and repeated 2-3 times. The results were averaged. As expected, we can conclude that as interfering conditions vary, so does the maximum reliable distance and to the lesser extent the maximum reliable angle. Table II summarizes the obtained data.

TABLE II
MAXIMUM DISTANCE AND MAXIMUM ANGLE FOR RELIABLE TRANSMISSION UNDER DIFFERENT CONDITIONS

| Condition | Maximum reliable distance [cm] | Maximum reliable angle [deg] |
|---|---|---|
| Declared | 370 | 30 |
| Measured w/o and with interfering light sources | | |
| No external light | 250 | 38 |
| Ambient light | 230-240 | 34 |
| Unmodulated LED | 100-220 | 36 |
| Direct sunlight | 60-180 | 26 |

*B. Testing with multiple luminaires*

The experiment was set up to verify simulation results. We used three lamps, placed at a distance of 16 cm from each other. We collected tag read-outs with the receiver moving directly under the lamps on a parallel plane distanced 30 cm from the luminaires (*d*). The horizontal distance corresponds to *x* from Fig. 7. Fig. 8 shows the obtained results – the resolution of tag readouts is compatible with the results of the simulation, and the error rate (number of bad or inconclusive tag readouts) was ~ 5%, with errors occurring in the transition area.

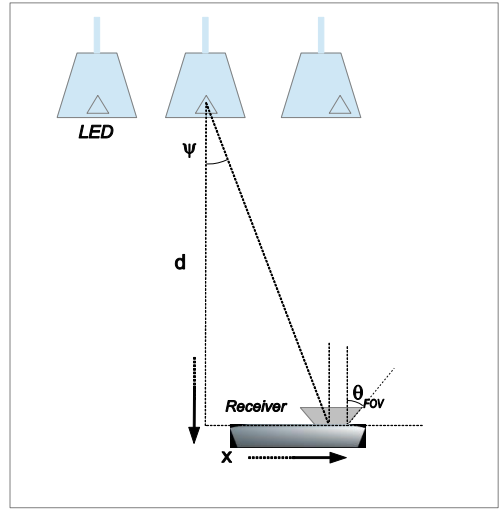

Fig. 7. Single and multiple transmitter experiment setup.

## VI. CONCLUSION

We tested a multi-tag LED-ID system both via numeric simulations and by means of an experiment. We have concluded that in a dense transmitter setup, i.e. with overlapping light cones, it is still possible to resolve transmitted digital tags, up to the point where light cones start to totally overlap. The methodology that we have presented should be useful for planning more complex LED-ID scenarios. It should also be helpful to the vendors of LED-ID hardware.

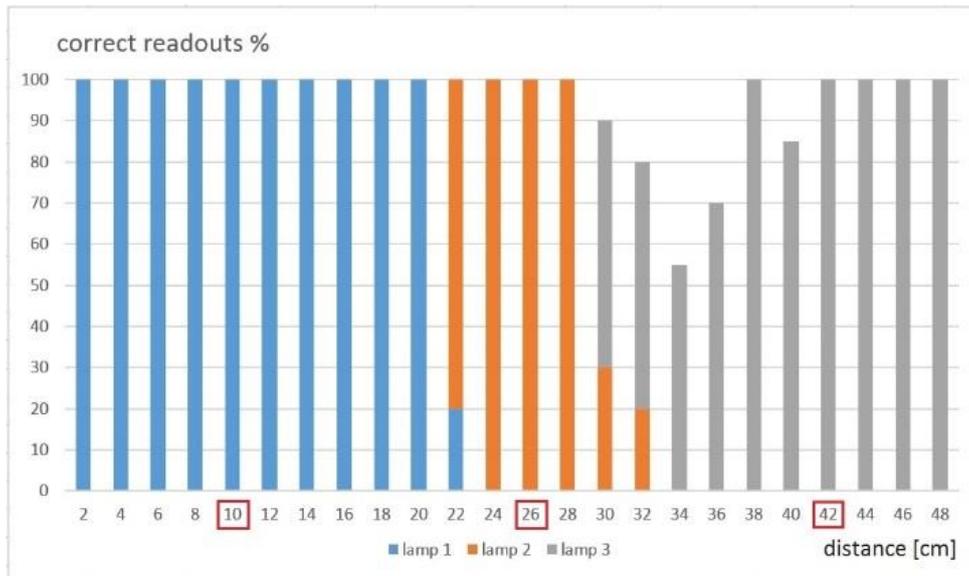

Fig. 8. Summary of tag readouts from experiment. Transmitters were placed at positions: 10, 26, 42 cm (marked as squares on the axis).

ACKNOWLEDGMENT

The authors would like to thank the anonymous reviewers for their helpful comments in the preparation of this article.